\documentclass{aa}  

\def\heao1{{\it HEAO-1\/}}

\def\lesssim{\mathrel{\hbox{\rlap{\hbox{\lower4pt\hbox{$\sim$}}}\hbox{$<$}}}}
\def\gtrsim{\mathrel{\hbox{\rlap{\hbox{\lower4pt\hbox{$\sim$}}}\hbox{$>$}}}}

\usepackage{rotating}
\usepackage{xcolor}

\DeclareGraphicsExtensions{.jpg,.png}

\begin{document}

\title{The ALPINE-ALMA [CII] Survey: Dust emission effective radius up to 3 kpc in the Early Universe}

\author{F. Pozzi\inst{1,2}, F. Calura\inst{2}, Q. D'Amato\inst{1,3}, M. Gavarente\inst{1}, M. Bethermin\inst{4,5}, M. Boquien\inst{6}, V. Casasola\inst{7}, A. Cimatti\inst{1}, R. Cochrane\inst{8,9} , M. Dessauges-Zavadsky\inst{10}, A.Enia\inst{2}, F. Esposito\inst{1}, A.L. Faisst\inst{11}, R. Gilli\inst{2}, M. Ginolfi\inst{12,3}, R. Gobat\inst{13}, C. Gruppioni\inst{2}, C. C. Hayward\inst{9}, E. Ibar\inst{14},  A.M. Koekemoer\inst{15}, B. C. Lemaux\inst{16,17}, G. E. Magdis\inst{18,19,20},   J. Molina\inst{14}, M. Talia\inst{1}, L. Vallini\inst{2}, D. Vergani\inst{2}, G. Zamorani\inst{2} }

{
\institute{Dipartimento di Fisica e Astronomia, Universit\'a di Bologna, via Gobetti 93/2, 40129, Bologna, Italy, email: f.pozzi@unibo.it
\and
INAF -- Osservatorio di Astrofisica e Scienza dello Spazio di
Bologna, via Gobetti 93/3, 40129, Bologna, Italy
\and
INAF -- Osservatorio Astrofisico di Arcetri
Firenze, via Largo Enrico Fermi 5, Firenze, Italy
\and 
Université de Strasbourg, CNRS, Observatoire astronomique 414
de Strasbourg, UMR 7550, 67000 Strasbourg, France
\and 
Aix Marseille Univ, CNRS, CNES, LAM, Marseille, France
\and
Universit\'e C\^{o}te d'Azur, Observatoire de la C\^{o}te d'Azur, CNRS, Laboratoire Lagrange, 06000, Nice, France
\and
INAF -- Istituto di Radioastronomia
Bologna, via Gobetti 111, 40129, Bologna, Italy
\and
Department of Astronomy, Columbia University, New York, NY 10027, USA
\and
Center for Computational Astrophysics, Flatiron Institute, 162 Fifth Avenue, New York, NY 10010, USA
\and 
Department of Astronomy, University of Geneva, Chemin Pegasi 51, 1290 Versoix, Switzerland
\and
Infrared Processing and Analysis Center, California Institute of
Technology, Pasadena, CA 91125, USA
\and
Dipartimento di Fisica e Astronomia, Università degli Studi di 470
Firenze, Via G. Sansone 1,I-50019, Sesto Fiorentino, Firenze, Italy 471
\and 
Instituto de Física, Pontificia Universidad Católica de Valparaíso, Casilla 4059, Valparaíso, Chile
\and
Instituto de Física y Astronomía, Universidad de Valparaíso, Avda. 
Gran Bretaña 1111, Valparaíso, Chile
\and
Space Telescope Science Institute, Baltimore, MD 21218, USA
\and
Department of Physics and Astronomy, University of California 454, Davis, One Shields Avenue, Davis, CA 95616, USA 455
\and
Cosmic Dawn Center (DAWN), Jagtvej 128, DK2200 Copenhagen 456
N, Denmark
\and
DTU-Space, Technical University of Denmark, Elektrovej 327, 458
DK2800 Kgs. Lyngby, Denmark 459
\and
Niels Bohr Institute, University of Copenhagen, Jagtvej 128, DK- 460
2200 Copenhagen N, Denmark
}

\authorrunning{F. Pozzi, F. Calura et al. }
\titlerunning{The ALPINE-ALMA survey: Dust mass budget in the early
  Universe}

\date{Received December 18, 2023; accepted March 20, 2024}

\abstract
{}
{Measurements of the size of dust continuum emission are an important tool for constraining the spatial extent of star formation and hence the build-up of stellar mass.
Compact dust emission has generally been observed at Cosmic Noon ($z\sim{2-3}$). However, at earlier epochs, toward the end of the Reionization ($z\sim{4-6}$), only the sizes of a handful of IR-bright galaxies have been measured. In this work, we derive the dust emission sizes of main-sequence galaxies at $z\sim5$ from the ALPINE survey.}
{We measure the dust effective radius $r_{e,FIR}$ in the uv-plane in Band 7 of ALMA for seven ALPINE galaxies with resolved emission and we compare it with rest-frame UV and [CII]158$\mu$m measurements. We study the $r_{e,FIR}-$L$_{IR}$ scaling relation by considering our dust size measurements and all the data in literature at $z\sim{4-6}$.  Finally, we compare our size measurements with predictions from simulations. }
{The dust emission in the selected ALPINE galaxies is rather extended ($r_{e,FIR}\sim{1.5-3}$ kpc), similar to [CII]158$\mu$m but a factor of $\sim$2 larger than the rest-frame UV emission. Putting together all the measurements at $z\sim5$, spanning 2 decades in luminosity from L$_{IR} \sim 10^{11}$ L$_\sun$ to L$_{IR}\sim 10^{13}$ L$_\sun$, the data highlight a steeply increasing trend of the $r_{e,FIR}-$L$_{IR}$ relation at L$_{IR}< 10^{12}$ L$_\sun$, followed by a downturn and a 
decreasing trend at brighter luminosities. Finally, simulations that
extend up to the stellar masses of the ALPINE galaxies considered in
the present work predict a sub-set of galaxies ($\sim25$ \%  at 10$^{10}$ M$_{\sun} <~ $M$_\star < 10^{11}$ M$_{\sun}$) with sizes as large as those measured.}
{}

{}

\maketitle

%

\section{Introduction}
\label{intro_sec}
The size of dust continuum emission in the far-IR (FIR) rest-frame
 regime gives key information on the spatial extension of the
dust-obscured star formation within galaxies, hence providing important
constraints on their stellar build-up. 

Nowadays, thanks to the Atacama Large Millimeter/Sub-millimeter array (ALMA), a significant number of studies
of the far-IR emission sizes have been performed at the Cosmic Noon ($z\sim2$) pointing towards compact dust emission (effective radius $r_{e,FIR} < 1-2$ kpc), typically more compact than the rest-frame optical/UV imaging (e.g. \citealt{2016ApJ...827L..32B}; {\citealt{2017ApJ...834..135T}, \citealt{2018MNRAS.476.3956T},  \citealt{2020A&A...643A..30F}). The compaction of the dust continuum sizes has been interpreted as a sign of dust-obscured build-up of a central
dense stellar component, either through the secular funneling of gas towards the center (e.g. \citealt{2013MNRAS.435..999D}) or through gas-rich mergers (e.g. \citealt{2021MNRAS.508.5217P}). These studies focus mainly on dusty, massive and FIR-bright galaxies while there are some hints that at
fainter luminosities dust emission is more extended (e.g. \citealt{2016ApJ...833...12R}, \citealt{2020MNRAS.499.5241C}). 

At very high redshift, measuring the dust
emission size is very challenging due to limited spatial resolution and sensitivity. Few studies are available for sources emerging shortly after the Epoch of Reionization ($4 < z < 6$). These studies include 
\cite{2022A&A...658A..43G}, who studied 6 sources in the GOODS-ALMA 2.0 survey at 1.1 mm; \cite{2022A&A...665A...3J} who considered  6 galaxies from the
super-deblended catalogues in COSMOS and GOODS-North ({\citealt{2018ApJ...864...56J}, {\citealt{2018ApJ...853..172L}) and 
\cite{2018ApJ...861..100C}, where the dust continuum sizes of 6 stacked luminous Sub-Millimeter Galaxies have been presented.
These works support the very compact dust continuum sizes found at lower redshifts, however the samples still rely on poor statistics and only the bright IR tail of galaxies has been probed (L$_{IR}(8-1000~{\mu}$m) > $10^{12}$ L$_\sun$).
An exception is the recent work from \cite{2022MNRAS.515.1751W}, which presented the properties of the ISM of 5 Lyman Break Galaxies (LBGs) at very high $z$ ($z\sim7$), characterised by moderate IR luminosity (L$_{IR} ~{\sim}~ 10^{11}$ L$_\sun$).

We aim to improve our understanding of the dust emission sizes at high $z$ ($4 < z < 6$) by taking advantage of the completed ALMA Large program to INvestigate [CII] at Early times (ALPINE, see \citealt{2020A&A...643A...1L}, \citealt{2020A&A...643A...2B}, \citealt{2020ApJS..247...61F}). 
The goal of this program was to observe the prominent [CII]
  158~${\mu}$m line and far-IR continuum emission for 118 UV-selected normal star-forming
  galaxies at $z\sim4.4-5.8$. 
  The present study increases the number of sources with measured dust continuum sizes at very high $z$ and, given the lower IR luminosity range spanned by the ALPINE galaxies, better constrains the $r_{e,FIR}$-L$_{IR}$ scaling relation. 
 
The paper is organised as follows. In Sect. 2 we describe our data analysis. We then present our result in Sect. 3. Finally, we discuss them and conclude in Sect. 4. Throughout the paper, we assume a flat $\Lambda$CDM cosmology with
$\Omega_{m}=0.3$, $\Omega_{\lambda}=0.7$ and $h_0=0.7$.

\section{Sample, observations and data analysis}
\label{sample_sec}


\begin{figure}
\center
\includegraphics[width=0.5\textwidth,angle=0]{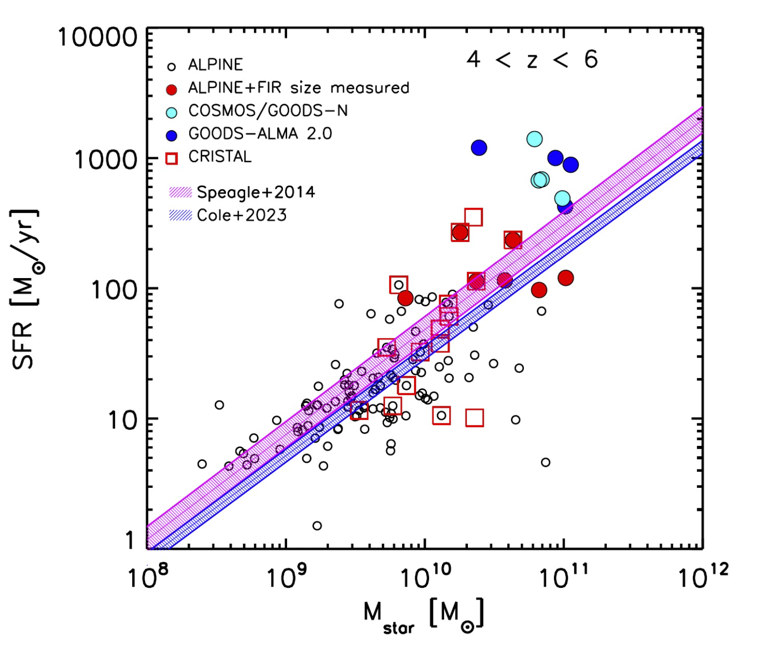}
\caption{SFR versus M$_{\star}$ for the ALPINE targets (red circles indicate galaxies with a FIR size measurement from the present work) and for our compilation of dust continuum measured sizes from literature in a similar redshift range $4<z<6$ (blue
  circles: COSMOS/GOODS-North galaxies from super-deblended catalogues
   (\citealt{2022A&A...665A...3J},  ; cyan circles: galaxies from the
   GOODS-ALMA 2.0 survey (\citealt{2022A&A...658A..43G}); red empty squares: galaxies from the CRISTAL survey (\citealt{2023arXiv231117671M}). For the CRISTAL sources not belonging to the ALPINE survey, we use the SFR and M$_{\star}$ values as reported in Table 1 from \cite{2023arXiv231117671M}. The violet shaded region  represents the 1$\sigma$ range of the MS relation found by \cite{2023arXiv231210152C} in the redshift range closer to the data ($4.5<z<5$) and obtained using the SFR values derived by averaging the star formation histories over 100 Myrs timescales (see Table 2 in \citealt{2023arXiv231210152C}); the pink shaded region represents the MS relation and its 1$\sigma$ dispersion found by \cite{2014ApJS..214...15S} and computed, for consistency, at $z=4.75$}.
\label{figure_ms}
\end{figure}

The parent sample of galaxies analysed in this paper is the subset of 23 objects with a $\sim$158~$\mu$m continuum detection in the ALPINE
survey (\citealt{2020A&A...643A...2B}). The ALPINE survey aimed
to study the $\sim$158~$\mu$m [CII] and the rest-frame FIR continuum
emission of 118 Main-Sequence (MS) rest-frame UV selected galaxies at
$z\sim4.4-5.9$, in the COSMOS (\citealt{2007ApJS..172....1S}) and ECDFS
(\citealt{2002ApJS..139..369G}) fields. 
For an overall description of the survey we refer to \citet{2020A&A...643A...1L}; for the
data reduction we refer to \cite{2020A&A...643A...2B}; for the description of the
ancillary spectra and photometric data and the physical properties
obtained from the UV-to-optical SED-fitting, we refer to
\cite{2020ApJS..247...61F}. The stellar masses (M$_\star$) and the star formation rates (SFR) of the
ALPINE galaxies are in the range M$_{\star}=10^{8.4}-10^{11}$
M$_{\odot}$ and $1.5-270$ M$_{\odot}$/yr, respectively. The stellar masses M$_\star$
 are from Table A1 of \cite{2020ApJS..247...61F} and derived from
 the SED-fitting; as SFR we consider SFR$_{UV+IR}$, i.e. the sum of
 the unobscured (SFR$_{UV}$) and obscured (SFR$_{IR}$)
 star formation. We derive SFR$_{UV}$  from the UV luminosity at $1500$ {\rm \AA} (\citealt{2020ApJS..247...61F}) uncorrected for dust, and  SFR$_{IR}$ from the IR luminosity reported in \cite{2020A&A...643A...2B}.
 
In Fig. \ref{figure_ms} we report the ALPINE galaxies on the
SFR$-M_{\star}$ plane. For the galaxies
not detected in continuum, following \cite{2020A&A...643A...5D}, we
consider only SFR$_{UV}$. As found by \cite{2020A&A...643A...3S}
by studying the UV-continuum slope, this assumption can
underestimate the total SFR by a factor of 2.  In Fig. \ref{figure_ms}, together with the ALPINE targets, we report the galaxies with measured continuum sizes in the redshift range $4 < z < 6$ (see \citealt{2022A&A...658A..43G}, \citealt{2022A&A...665A...3J}). Moreover, as a reference, we report the MS relation from  \cite{2023arXiv231210152C}, recently derived using JWST data from Cosmic Evolution Early Release Science (CEERS, \citealt{2022ApJ...940L..55F}, \citealt{2023ApJ...946L..12B}) and consistent, within 1${\sigma}$ with the pre$-$JWST derivation from \cite{2014ApJS..214...15S}. 
Over the 23 ALPINE continuum sources, 3 galaxies are classified in the continuum images as multi-components objects (VC$\_$5101209780,VE$\_$530029038, DC$\_$881725, see \citealt{2020A&A...643A...2B} ). These systems can indicate the presence of a merger or a very patchy and disturbed galaxy where can be quite difficult to constrain a center of the system and the determination of the size. 
For this reason, we choose to remove these galaxies from our
analysis. Therefore, our sample consists of 20 ALPINE
galaxies. Among these sources, 18 over 20 are detected in [CII], while
for 2 sources only upper limits in [CII] are available.


\begin{table*}
\centering
\begin{tabular}{lccccccc} \\\hline\hline
Name                          &  $z$        &$r_{e,FIR}$          &$r_{e,[CII]}$           &        $r_{e,UV}$    &            $logM_{\star}$ &      $SFR$ & $S/N$\\
                               &            &[kpc]              &      [kpc]           &      [kpc]         &[log$_{10}$(M/M$_{\odot}$)] & [M$_{\odot}$/yr]  \\\hline
  CG$\_$19              &  4.500      &  $<$1.40         &  \ldots    &   \ldots     &      9.83$^{+0.07}_{-0.05}$       &  66.5$^{+13.5}_{-11.0}$ &5.4\\
DC$\_$396844     &  4.540     & 2.13$\pm$1.10    &     2.56$\pm$0.33  & 0.58$\pm$0.20 &     9.86$^{+0.14}_{-0.19}$       &  83.9$^{+16.6}_{13.8}$  &5.6 \\
DC$\_$488399     &  5.678     &  $<$      1.17    &   1.32$\pm$0.16   &   \ldots    &      10.20$^{+0.13}_{-0.15}$       &   89.8$^{+14.9}_{-13.1}$ &9.3\\
DC$\_$494057     &  5.540     &  $<$      1.16    &     2.48$\pm$0.25 &  0.59$\pm$0.17 &   10.15$^{+0.13}_{-0.15}$       &  77.8$^{+16.1}_{-13.8}$ &7.1\\
DC$\_$552206     &  5.514     & 3.08$\pm$ 1.25    &  \ldots           &       \ldots    &  10.58$^{+0.14}_{-0.16}$       & 114.9$^{+29.6}_{-23.3}$ &4.5\\
DC$\_$683613$^{a}$    &  5.536     &  $<$      1.65 & 1.82$\pm$0.33 &  0.57$\pm$0.24  &     10.17$^{+0.14}_{-0.15}$      &  75.6$^{+18.6}_{-14.9}|$ &5.1\\
DC$\_$818760$^{a}$     &  4.554     & 2.70$\pm$ 0.55  &  2.59$\pm$0.16 &  1.07$\pm$0.21$^{b}$  & 10.63$^{+0.11}_{-0.10}$ & 235.1$^{+28.6}_{-25.5}$ &9.5\\
DC$\_$848185$^{a}$     &  5.284     & 1.90$\pm$ 0.30  &    \ldots      &  \ldots      &     10.37$^{+0.08}_{-0.19}$      & 113.0$^{+16.0}_{-14.3}$ &7.1\\
DC$\_$873756$^{a}$     &  4.548     & 1.58$\pm$ 0.20    &  2.36$\pm$0.11 & 1.08$\pm$0.43    &     10.25$^{+0.08}_{-0.10}$ & 268.7$^{+16.7}_{-15.5}$ &21.7\\
VC$\_$5100822662$^{a}$  &  4.523    &  $<$1.16    &   2.59$\pm$0.37 &  1.32$\pm$0.33     &     10.17$^{+0.13}_{-0.14}$   &  60.8$^{+10.5}_{-9.0}$   &6.5\\
VC$\_$5101218326     &  4.568     & 1.83$\pm$0.40     &   2.37$\pm$0.15 &  1.46$\pm$0.32     &     11.01$^{+0.07}_{-0.05}$& 120.0$^{+18.0}_{-15.6}$ &6.5\\
VC$\_$5180966608     &  4.529     & 2.60$\pm$0.60     &     no data  &  0.59$\pm$0.24$^{b}$  &     10.82$^{+0.12}_{-0.13}$ &  96.9$^{+19.5}_{-16.0}$&5.7 \\\hline
\end{tabular}
\caption{Column (1): ALPINE source name. We only list the 12
  sources with a $S/N$ of the continuum $>4.5$ (see \citealt{2020A&A...643A...2B}) that are considered in our analysis. The three sources classified in
  \cite{2020A&A...643A...2B} as multi-components objects have not been
considered. The (a) symbol indicates galaxies in common with he CRISTAL sample (\citealt{2023arXiv231117671M}). Column (2): spectroscopic redshift from the [CII] line
emission. Column (3): dust continumm effective radius (see text for details). Column (4): [CII] line emission effective radius
(\citealt{2020ApJ...900....1F}). Column (5): rest-frame UV effective
radius (\citealt{2020ApJ...900....1F}). The (b) symbol indicates that
the HST/F160W map has been considered, otherwise the radius from the
HST/F814W maps have been reported. Column (6), Column (7): stellar
mass and SFR (see text for details).  Column (8): Continuum S/N at 158~$\mu$m from \cite{2020A&A...643A...2B}.}
\label{table1}
\end{table*}


\begin{figure*}
\begin{center}
\includegraphics[width=1.\textwidth,angle=0]{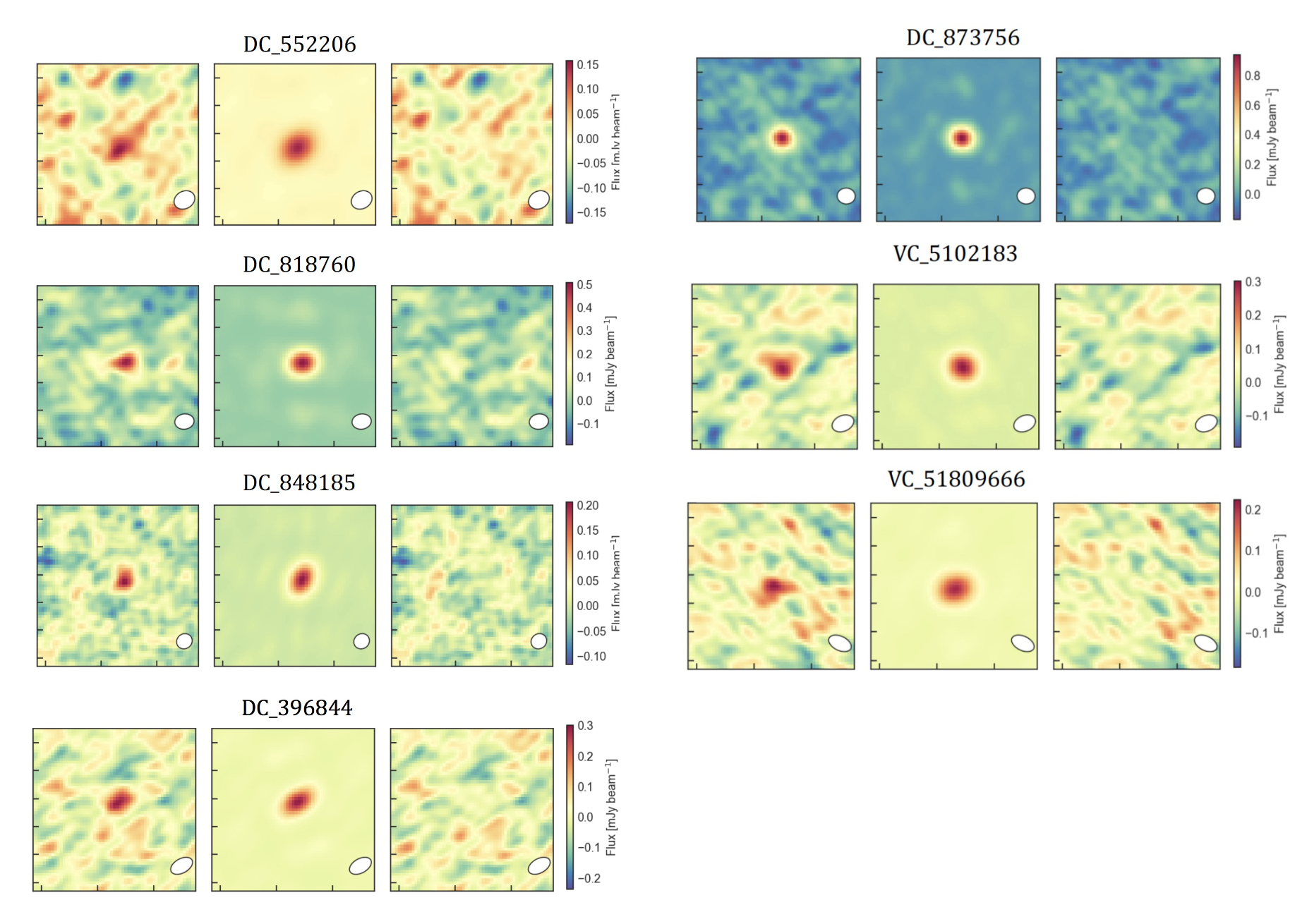}
\caption{Rest-frame FIR size measurements for the 7 resolved ALPINE sources using the CASA task UVMODELFIT. For every source, the three panels are, from left to right, the observed, modelled, and residual maps, respectively, all with a $10^{\prime\prime}{\times}10^{\prime\prime}$ size. }
\label{figure_images}
\end{center}
\end{figure*}


\begin{figure*}
\center
\includegraphics[width=1\textwidth,angle=0]{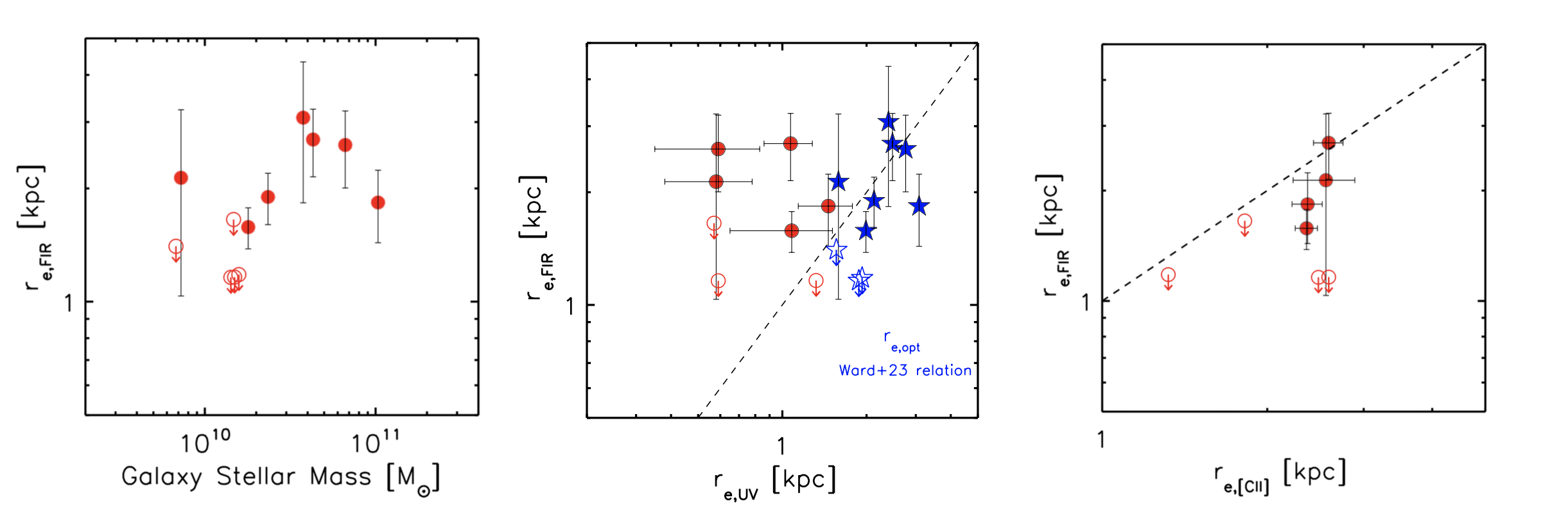}
\caption{Left: Effective radius $r_{e,FIR}$ vs. stellar mass M$_\star$. Centre: $r_{e,FIR}$ vs. $r_{e,UV}$. The blue symbols represent the $r_{e,FIR}$ vs. the optical radius $r_{e,opt}$, with $r_{e,opt}$ from the empirical JWST mass$-$size relation (\citealt{2023arXiv231102162W}) given the stellar masses of our sources. Right: $r_{e,FIR}$ vs. $r_{e,[CII]}$. The ALPINE galaxies continuum detected at S/N > 4.5 have been considered for the size analysis. The unresolved sources are reported as empty circles. Both $r_{e,UV}$ and
$r_{e,[CII]}$ are from \cite{2020ApJ...900....1F} and
only the sources with a reliable measurements (flag=0) have been considered.}
\label{figure_sizes}
\end{figure*}


\subsection{Size measurement}
\label{size_sec}

In order to measure the continuum sizes of the targets, we adopted
the procedure outlined in \cite{2020ApJ...900....1F} to measure the
sizes of the [CII] emission of the ALPINE targets. First of all, we
remove from the visibility data the channels containing the [CII]
emission. We then produce the continuum maps with the CASA\footnote{Common Astronomy Software Application. \url{https://casa.nrao.edu/}}(\citealt{2022PASP..134k4501C}) task
TCLEAN and we run in CASA the task
IMFIT to achieve a first guess of the target's properties (positions,
fluxes and sizes). These outputs, obtained in the image-plane, are used
as first guess for the proper size measurements in the uv-plane, where we assume a 2-dimensional Gaussian model for the
intensity profile and we measure the sizes with the CASA task
UVMODELFIT. The latter returns the de-convolved FWHM
along the major and minor axis (FWHM$_{maj}$, FWHM$_{min}$) that we circularise as 
FWHM$_{circ}=\sqrt{\text{FWHM}_{min}{\times}\text{FWHM}_{maj}}$). The
effective radius $r_{e}$, which represents the radius that encloses
half of the total light, is then derived as $r_{e}=$FWHM$_{circ}$/2 (see \citealt{2010MNRAS.404..458V} for
a Gaussian profile corresponding to a S\'ersic index $n$=0.5). \cite{2020ApJ...900....1F} describe how a Gaussian profile return an effective radius $r_{eff}$
consistent at a 5$\%$ level with a more reliable exponential shape (corresponding to a S\'ersic index $n$=1).
We applied the described procedure to sources detected in the continuum at $S/N>4.5$ to obtain reliable measurements. Given this cut in $S/N$, our final sample is composed of 12 galaxies.  All but one of the 12 galaxies have  continuum $S/N>4.5$ (see Table \ref{table1}), and this gives us confidence of meaningful size measurements (\citealt{2020ApJ...900....1F}).

In Table \ref{table1} we report our size measurements. For 7 sources
the fitting procedure returns a de-convolved FWHM along the minor and major axes, and we obtain a circularized effective radius ($r_{e,FIR}$). 
We also visually check the observed, model and residual maps in the image plane to exclude multi-components below the scale of the beam (see Fig. \ref{figure_images}). We note that for source DC$\_$818760, constituted by three galaxies (see \citealt{2020MNRAS.491L..18J} and {\citealt{2023arXiv231111493D}), we measure and report the size of the central galaxy. For the remaining sources, since the fitting procedure reports de-convolved FWHM uncertainties larger or similar to the de-convolved FWHM measures, we estimate the size upper limit from the formula by \cite{2012A&A...541A.135M} considering for the likelihood $\lambda_c$ a value equal to 9 corresponding to a 3$\sigma$ cut-off. 
We note that 3 resolved sources (DC$\_$818760, DC$\_$848185, DC$\_$873756) are in common with the recent results from the CRISTAL survey (\citealt{2023arXiv231117671M}) and the size measurements are consistent within 1$\sigma$. Among the 5 unresolved sources, 2 sources (DC$\_$683613, VC$\_$5100822662) have a measured size in the CRISTAL sample consistent with our upper limits. The sources in common are highlighted in Table \ref{table1}. In Table \ref{table1} along with $r_{e,FIR}$, we report 
the circularized effective radius of the [CII] line emission ($r_{e,[CII]}$) and of the
rest-frame UV emission ($r_{e,UV}$). These radii were derived by
\cite{2020ApJ...900....1F} as follows: the [CII] radius from the ALMA ALPINE data
using the CASA UVMODEL task described above; the rest-frame UV
emission from the F814W HST map (\citealt{2007ApJS..172..196K}) and the F160W HST maps (\citealt{2011ApJS..197...36K}) using the GALFIT task (\citealt{2010AJ....139.2097P}). We only consider reliable size measurements (flag=0).  Finally, in Table \ref{table1} we give the rest$-$frame optical radius $r_{e,opt}$ derived from the empirical stellar mass$-$size relation found recently by the JWST CEERS survey (\citealt{2023arXiv231102162W}), given the stellar masses of our galaxies. We consider the mass$-$size relation in the redshift range closer to our data ($3.5<z<5$, see Table 1 in \citealt{2023arXiv231102162W}).


\section{Results}
\label{result_sec}

\subsection{FIR, [CII] and UV sizes}
\label{result_sec_sizes}

In Fig. \ref{figure_sizes} we report the $r_{e,FIR}$ of the 12 ALPINE targets considered in this work with a $S/N>4.5$ continuum detection (upper limits for the unresolved sources are displayed as open circles). In the left panel $r_{e,FIR}$ is presented as a
function of M$_{\star}$, in the middle panel as a function of the
rest-frame UV sizes $r_{e,UV}$ and of the optical radius $r_{e,opt}$
as derived by the JWST stellar mass$-$size relation
(\citealt{2023arXiv231102162W}),  and in the right panel as a function
of the [CII] sizes $r_{e,[CII]}$. 

We find the median value to be $r_{e,FIR}=(2.13\pm0.26)$ kpc, and the ratios
$r_{e,UV}/r_{e,FIR}=(0.39\pm0.15)$ and
$r_{e,[CII]}/r_{e,FIR}=(1.29\pm0.14)$, calculated considering only the 7
rest-frame FIR resolved sources and the 5(4)
galaxies with measured $r_{e,UV}(r_{e,[CII]}$.  The median ratio of
the dust continuum sizes over the optical ones derived from the
mass$-$size relation is $r_{e,opt}/r_{e,FIR}=1.10\pm0.26$. The $r_{e,FIR}$ sizes are larger than those found for brighter FIR sources at similar redshifts (see the review from \citealt{2020RSOS....700556H} and
 references therein). They are slightly smaller than the [CII] sizes, tracing both the atomic and the
 molecular gas, and significantly larger (a factor of 2.5) than
 the rest-frame UV sizes, opposed to what has been found by other authors ($r_{e,HST}/r_{e,ALMA}=1.6$ for galaxies at $1<z<6$ taken from the ALMA archive by  \citealt{2017ApJ...850...83F};  $r_{e,HST}/r_{e,ALMA}=2.3$ for a compilation of massive galaxies at $z=2-4$ in the GOODS-S by \citealt{2020A&A...643A..30F}) and in agreement with results from CRISTAL survey (\citealt{2023arXiv231117671M}). Moreover, our $r_{e,UV}/r_{e,FIR}$ are significant smaller (up to a factor of 5-10) than the values predicted by the hydrodynamical cosmological simulations from \cite{2022MNRAS.510.3321P} and extrapolated at the  stellar masses of our galaxies. This suggests that the model from \cite{2022MNRAS.510.3321P}  could predict too compact FIR dust emission extension.  Finally, we find that the dust continuum sizes are very close to the optical ones, as predicted from the JWST mass$-$size relation, pointing towards a quite extended, disk-like star-formation region traced by the IR emission, very similar to the region traced by the rest-frame optical emission. 
  We will discuss in Sect. \ref{result_sec_sizes_fir_mstar} our findings in comparison with
 other results from literature and predictions from models.}


\subsection{FIR sizes versus  IR luminosity and stellar masses}
\label{result_sec_sizes_fir_mstar}
The study of scaling relations from infrared data stands as a cornerstone in astrophysics, offering a comprehensive perspective on the properties of galaxies related to their coldest components, 
i.e. neutral gas, molecular gas and dust (\citealt{2017MNRAS.465...54C,2018ApJ...861...95H,2020A&A...633A.100C,2020MNRAS.493.3580P}).
A poorly explored, yet crucial quantity, is the size of dust emission in high-redshift galaxies. 
In this work, we aim to leverage this essential parameter to investigate pivotal scaling relations, i.e. the ones between size vs luminosity and size vs. stellar mass.

In Fig. \ref{figure_sizes_lir}, we report our measures of $r_{e,FIR}$ as a
function of the IR luminosity L$_{IR}$. Here, L$_{IR}$
has been computed  from the rest-frame 158$\mu$m 
emission assuming the \cite{2017A&A...607A..89B} SED. 
  
Together with our measurements, we report all the available
rest-frame FIR size measurements from the literature at similar redshifts ($4 < z <  6$). To avoid bias introduced by the radial gradient in dust temperature (e.g. \citealt{2018ApJ...863...56C}; \citealt{2019MNRAS.488.1779C}), we consider galaxies observed at nearly the same wavelengths of the ALPINE galaxies ($0.8-1$ mm): this allows us to sample the rest-frame continuum emission at the same frequencies, given the same redshift range.  
In blue symbols we show the measurements from the GOODS-ALMA 2.0 survey
(\citealt{2022A&A...658A..43G}). The GOODS-ALMA 2.0 survey is an ALMA blind survey at 1.1 mm
covering an area of 72.42 arcmin$^2$. The rest-frame FIR sizes
were derived with a circular Gaussian model fit in the uv-plane, and combining
the low and high-resolution observations, leading to an average resolution of
0.447$^{\prime\prime}$ ${\times}$ 0.418$^{\prime\prime}$. Among the 88 blind detected sources in the GOODS-ALMA 2.0 survey, 4 sources satisfy our adopted selection criteria ($4 < z <  6$), with the redshift $z$ estimated from photometry (\citealt{2022A&A...658A..43G}).
In cyan symbols we report the dust continuum sizes of 6 galaxies from the super-deblended COSMOS and GOODS-North catalogues, selected for their very high photometric redshift ($z$ > 6, \citealt{2022A&A...665A...3J}). Thanks to the detection of [CI](1-0) and CO transitions, the authors were able to measure the spectroscopic redshifts and among the 6 sources, 4 sources
satisfy our criteria (4 < $z$ <
6). The dust continuum sizes are derived from 870 $\mu$m and 1 mm ALMA data.


\begin{figure}
\centering
\includegraphics[width=0.5\textwidth,angle=0]{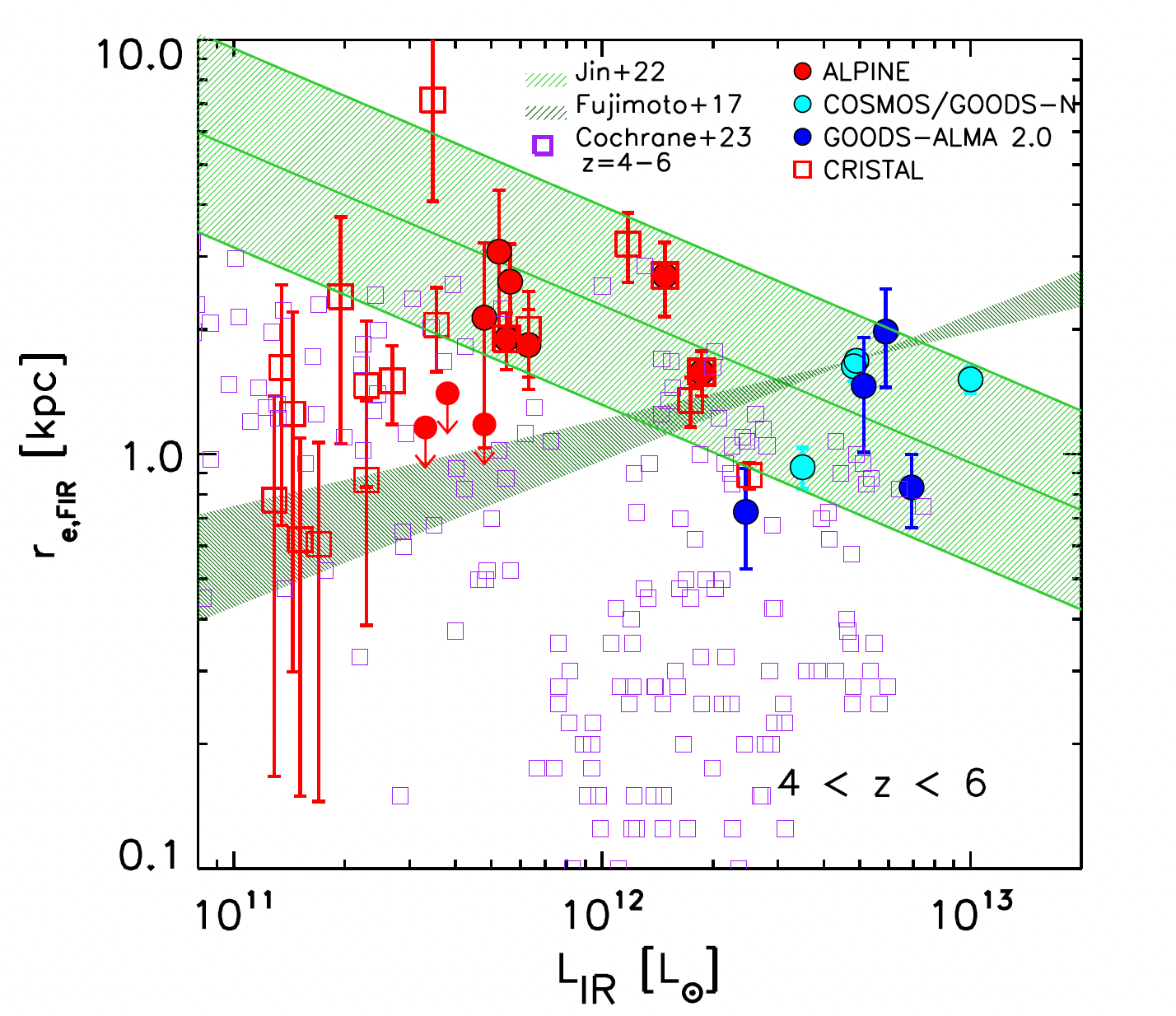}
\caption{Effective radius $r_{e,FIR}$ vs. IR luminosity L$_{IR}$ at $4<z<6$ for this
sample and from the literature. Red circles: ALPINE survey; blue circles: GOODS-ALMA 2.0 survey (\citealt{2022A&A...658A..43G}); cyan circles:
super-deblended catalogs in the COSMOS and GOODS-North fields
(\citealt{2022A&A...665A...3J}); red open squares: CRISTAL survey (\citealt{2023arXiv231117671M}).The dark-green shaded region  represents the 1$\sigma$ range of the relation found by \cite{2017ApJ...850...83F}; the green lines and the shaded region
represent the relation and its 1$\sigma$ dispersion found by \cite{2022A&A...665A...3J}. Pink empty squares are simulation results 
from \cite{2023MNRAS.518.5522C} in the range $4 <z < 6$.}
\label{figure_sizes_lir}
\end{figure}


We also report as red open squares the recent measurements from \cite{2023arXiv231117671M} from the CRISTAL survey  (see Herrera-Camus et al. in prep. for the survey paper). The CRISTAL survey is composed by 24 target galaxies, of which 20 from the ALPINE survey (5 in common with the sample analysed in the present work, see Table \ref{table1}). This survey has been carried out at the same wavelengths of the
ALPINE survey (Band 7: $0.8-1$ mm), but at higher spatial resolution ($\sim0.3^{\prime\prime}$).   
As clear from Fig. \ref{figure_sizes_lir}, the CRISTAL sample probes the faintest galaxies studied so far, 
with FIR luminosities down to L$_{IR}\sim 10^{11} L_{\odot}$, while our data probe a luminosity domain scarcely populated by the dataset of \cite{2023arXiv231117671M}; in this regard, 
the CRISTAL sample and the ALPINE dust continuum resolved sources from the present work are complementary. The analysis of a dataset that probes bright galaxies alongside another one probing fainter counterparts, all at similar redshifts,  provides us with a comprehensive understanding, allowing us to characterise different trends of the $r_{e,FIR}-$L$_{IR}$  relation in distinct regimes of luminosity and size. 

In Fig. \ref{figure_sizes_lir}, we report
for comparison 2 relations: the one found by
\cite{2017ApJ...850...83F} (dark-green shaded region, representing the 1$\sigma$ dispersion) and the one of  \cite{2022A&A...665A...3J} (green line with its 1$\sigma$ dispersion
displayed as shaded region). The result of \cite{2017ApJ...850...83F} is
based on a large sample of galaxies from the ALMA
archive. Considering the sources all together ($0 < z < 6$), these authors
find a positive correlation between $r_{e,FIR}$ and L$_{IR}$ ($r_{e,FIR}~{\propto}~L_{IR}^\alpha$ with
$\alpha=0.28\pm0.07$). 

This correlation is confirmed by the
authors also in the
highest redshift range considered (2 $< z
< 4$). \cite{2017ApJ...850...83F}  discussed that the origin of the $r_{e,FIR}-$L$_{IR}$ relation could be related to the formation of stellar
disks, since the IR slope is similar to the
slope observed in the UV band, the latter explained by the predictions of disk formation models
(i.e. \citealt{2014ApJ...788...28V}). On the other side, the relation from
\cite{2022A&A...665A...3J}  is based on a different compilation of measurements
at $z > 1$: the 6 galaxies analysed in their work, the galaxies from the GOODS-ALMA
(\citealt{2020A&A...643A..30F}; \citealt{2022A&A...658A..43G}) and a
sample of MS galaxies at $z=1-2$
(\citealt{2020ApJ...890...24V}). Opposite to \cite{2017ApJ...850...83F}, in the latter work the authors find an
anti-correlation between the dust size and L$_{IR}$ (log(FWHM
size/kpc)= $-$ 0.38 ${\times}$ log (L$_{IR}/10^{10}$ L$_{\sun}$)+1.42),  indicating that galaxies with higher luminosities tend to have a
  more compact dust morphology. 

 In Fig. \ref{figure_sizes_lir} we also show as pink empty squares the predictions from high-resolution, cosmological zoom-in simulations in a similar redshift range as sampled by the data ($4 <z<6$). These predictions are drawn from the FIRE (\citealt{2018MNRAS.480..800H}) suite described in \cite{2023MNRAS.518.5522C} and were generated using the radiative transfer methods described by \cite{2019MNRAS.488.1779C}, but extending the sample to lower FIR luminosities. To properly compare the model predictions to the data, we consider the observed-frame 850 $\mu$m emission sizes.

Concerning the data, a positive correlation between $r_{e,FIR}$ and L$_{IR}$ is supported at faint luminosities 
by the ALMA-CRISTAL data.
On the other hand, across a luminosity range wider than 1 dex and
extending from L$_{IR}\sim 10^{11.5}$ L$_\sun$ to L$_{IR}=2{\times} 10^{13}$ L$_\sun$, our measurements, together with the \cite{2022A&A...665A...3J} and \cite{2022A&A...659A.196G} data, support an anti-correlation between $r_{e,FIR}$ and L$_{IR}$ as found by  \cite{2022A&A...665A...3J}, at odd with the results from  \cite{2017ApJ...850...83F}. We are aware of the small number of galaxies with a measured dust continuum size at high-$z$ ($>4$) and high-L$_{IR}$ ($>10^{11.5}$ L$_\sun$); anyway, we caution the reader of possible observational biases which may affect the \cite{2017ApJ...850...83F} relation, given the sample inhomogeneity in terms of spatial resolution and sensitivity and the use of photometric redshift for deriving sizes and L$_{IR}$. Moreover, both the \cite{2022A&A...665A...3J} and \cite{2017ApJ...850...83F} relations are obtained from samples including galaxies at lower $z$ than the range considered in this work ($4<z<6$), and are both derived for IR-bright galaxies (L$_{IR}> 10^{11.4}$ and $>10^{12} $L$_\sun$, respectively).

Altogether, our results might suggest a variable trend of FIR luminosity as a function of size, that ranges from an increase for faint objects, a downturn at some characteristic luminosity of the order of $10^{12}$ L$_\sun$, followed by a decreasing trend at brighter luminosities. A similar trend is shown also by the compilation of measures of \cite{2023arXiv231117671M}, across a wider redshift range ($2\le z \le 6$).  The predictions from the zoom-in simulations (\citealt{2023MNRAS.518.5522C}) cover a broader range in dust continuum sizes than the data but, as the measurements, they present a characteristic IR luminosity around $10^{12}$ L$_\sun$, above which the sizes decrease. Understanding the physical reason of this downturn and accounting for the characteristic luminosity value at which it occurs will be a major challenge for galaxy formation models.


\begin{figure}
\centering
\includegraphics[width=0.52\textwidth,angle=0]{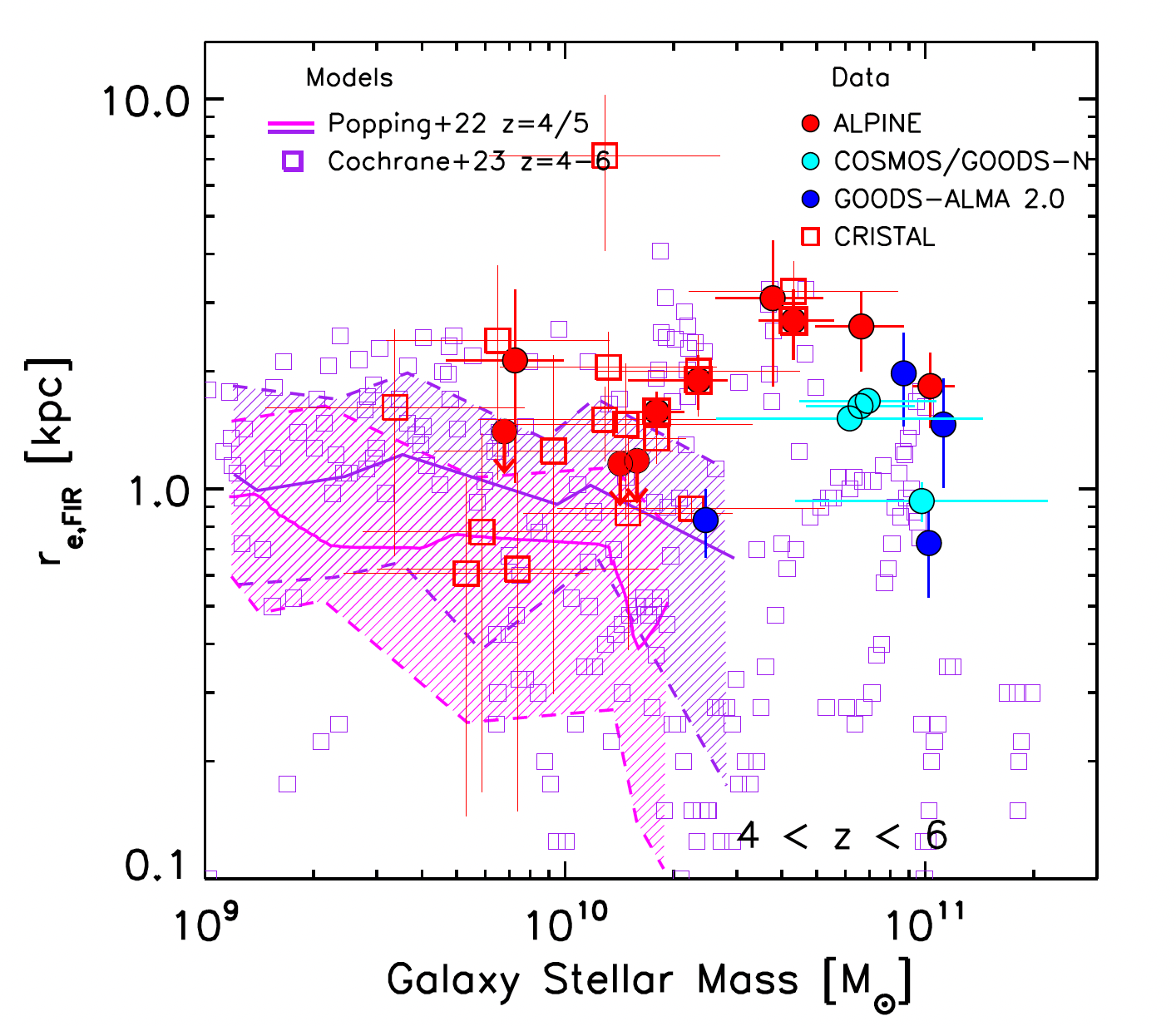}
\caption{Effective radius $r_{e,FIR}$ vs. stellar mass M$_{\star}$ at $4<z<6$ for this
sample and from the literature. Symbols as in Fig. \ref{figure_sizes_lir}. Magenta and violet lines (with shaded
regions marking the 1$\sigma$ scatter) represent the 
predictions from \cite{2022MNRAS.510.3321P} for $z=4$ and
for $z=5$ galaxies, respectively; pink empty squares are the
predictions from \cite{2023MNRAS.518.5522C} in the range $4 <z < 6$.}
\label{figure_sizes_mstar}
\end{figure}


In Fig. \ref{figure_sizes_mstar} we report our $r_{e,FIR}$ as a
  function of the stellar mass M$_\star$. As in Fig. \ref{figure_sizes_lir}, along with our measurements, we show the
  other samples with dust continuum sizes measured in the literature at $z>
  4$, from the GOODS-ALMA 2.0 survey
(\citealt{2022A&A...658A..43G}), from the super deblended catalogues in the COSMOS and GOODS-North fields
(\citealt{2022A&A...665A...3J}) and from the CRISTAL survey (\citealt{2023arXiv231117671M}).

In Fig. \ref{figure_sizes_mstar} together with the predictions from 
\cite{2023MNRAS.518.5522C} (pink empty squares for galaxies in the
range 4 $<z<6$) we also report the predictions from
\cite{2022MNRAS.510.3321P} (magenta and violet lines with shaded
regions marking the 1$\sigma$ scatter, for $z=4$ and $z=5$ galaxies, respectively). The predictions from
\cite{2022MNRAS.510.3321P} are derived from the TNG50 simulations
(\citealt{2019MNRAS.490.3234N}; \citealt{2019MNRAS.490.3196P}), the
highest resolution implementation of the magnetohydrodynamical
cosmological simulation IllustrisTNG (see \citealt{2018MNRAS.475..676S}; \citealt{2018MNRAS.480.5113M}).
Also for these predictions, we consider the observed-frame 850 $\mu$m emission sizes. Results for galaxy stellar masses up to
${\sim}10^{10.4}$ and ${\sim}10^{11.4}$ M$_{\sun}$ have been reported
in the \cite{2022MNRAS.510.3321P}  and \cite{2019MNRAS.488.1779C}
simulations, respectively. The stellar mass range of our galaxies is not sampled in the volume of the \cite{2022MNRAS.510.3321P} simulation. 
At masses $<10^{10}$ M$_{\sun}$, the model from  \cite{2022MNRAS.510.3321P} predicts compact dust emission sizes ($<~1-2$
kpc), similar to those observed in some CRISTAL galaxies (\citealt{2023arXiv231117671M}). 

At masses $>10^{10}$ M$_{\sun}$, 
a subset of model sources  ({$\sim 25\%$ in the range 10$^{10}$ M$_{\sun} < ~$M$_\star < 10^{11}$ M$_{\sun}$) from \cite{2023MNRAS.518.5522C} are predicted with large sizes, in the range of $1.5-4$ kpc, populating the upper envelope of the distribution and in agreement with the
ALPINE galaxies analysed in the present work.}

\section{Discussion and Summary}

Our study of the main-sequence ALPINE galaxies at $z\sim5$ shows effective radius $r_{e,FIR}$ of $1.5-3$ kpc, a factor of two larger than what is observed in brighter IR sources at similar $z$. We confirm, in fact, the anti-correlation found by \cite{2022A&A...665A...3J} between the dust size and the IR luminosity in the L$_{IR}$ range between  ${\sim}10^{11.5}$ L$_\sun$ and ${\sim}2{\times} 10^{13}$ L$_\sun$. A possible explanation for the anti-correlation 
is ascribed to galaxies caught at different phases of their star formation history, all within the scatter of the main-sequence (e.g. \citealt{2021MNRAS.503.4878S}; \citealt{2022A&A...659A.196G}; \citealt{2024MNRAS.527...10V}). In this picture, the ALPINE galaxies are characterised by a mild SFR, a more extended (possibly disk-like) star formation region, while IR brighter galaxies are characterised by
an enhanced SFR, short depletion time, and high dust temperatures. To validate this scenario, we consider the depletion times ($t_{depl}$) and the gas fractions ($f_{gas}=\frac{M_{gas}}{M_\star+M_{gas}}$) for the ALPINE galaxies with a measured dust size, derived by  \cite{2020A&A...643A...5D} from the [CII] luminosities: $t_{depl}=(456\pm{57})$ Myrs and
$f_{gas}=0.6\pm{0.2}$. These values are in agreement with those derived from \cite{2022A&A...659A.196G} for normal star-forming galaxies with an extended star-forming region, as opposite to the typical values of starburst galaxies, characterised by a
more compact emission, shorter $t_{depl}$ (of the order of 10-100 Myrs, see also \citealt{2015ApJ...812L..23S}, \citealt{2023ApJ...943...82S}) and lower gas fraction (in the range 0.25-0.5, see Fig. 3 in \citealt{2022A&A...659A.196G}). Our finding is in line with the recent result obtained by \cite{2023A&A...680L...8B} on the Kennicutt-Schmidt (KS) relation at $z\sim4.5$: they analysed 4 galaxies (3 in common with the present sample) at higher spatial resolution ($0.2^{\prime\prime}-0.3^{\prime\prime}$), showing how MS galaxies have significantly lower $\Sigma_{SFR}$, at a given $\Sigma_{gas}$, in comparison to high-$z$ starburst galaxies. 

On the other hand, in fainter galaxies, the new CRISTAL sample (\citealt{2023arXiv231117671M}) supports a positive correlation between IR size and IR luminosity. 
Altogether, the data samples considered in this work, spanning 2 dex in IR luminosity, highlight a variable trend in the $r_{e,FIR}-$L$_{IR}$ relation, ranging from 
steeply increasing behaviour at L$_{IR}< 10^{12}$ L$_\sun$, followed by a downturn and a 
decreasing trend at brighter luminosities. 

Finally, we compare our dust emission sizes with the cosmological simulations of \cite{2023MNRAS.518.5522C} and \cite{2022MNRAS.510.3321P}. Their results suggest smaller dust continuum sizes than our measurements but in \cite{2023MNRAS.518.5522C}, the only simulation that extends up to our stellar masses, a tail of sizes as large as those observed in the ALPINE galaxies is predicted. 

 Future ALMA observations at higher resolution and in different bands will strengthen this result by enlarging the sample of galaxies with a measured size and by allowing the estimate of another key parameter: the dust temperature.
 Besides being crucial in the estimate of the dust mass (e.g. \citealt{2013A&A...552A..89B}; \citealt{2021A&A...653A..84P}), this quantity is critically sensitive to the 
optical thickness of the dust-emitting region (e.g. \citealt{2022A&A...665A...3J}) and to the source physical area (hence its size, \citealt{2022A&A...666A..17G}). 
Further elucidating  the scaling relations presented here by means of improved observations and understanding the 
underlying physical mechanisms are important avenues for future investigation, and will provide fundamental insights in future galaxy evolution studies.

\begin{acknowledgements}
This paper is dedicated to the memory of Olivier Le F\`evre, PI of the 
ALPINE survey. We thank the referee for their insightful comments. We acknowledge support from grant PRIN MIUR
2017- 20173ML3WW$\_$001. FP, VC, FE and FC acknowledge funding from the INAF Mini Grant 2022 program “Face-to-Face with the Local Universe: ISM’s Empowerment (LOCAL)”. FC acknowledges support from the INAF
main-stream (1.05.01.86.31). EI acknowledges funding by ANID FONDECYT Regular 1221846.GEM acknowledges the Villum Fonden research grant 13160 “Gas to stars, stars to dust: tracing star formation across cosmic time,” grant 37440, “The Hidden Cosmos,” and the Cosmic Dawn Center of Excellence funded by the Danish National Research Foundation under the grant No. 140. MB acknowledges support from the French government through the France 2030 investment plan managed by the National Research Agency (ANR), as part of the Initiative of Excellence of Universit\'e C\^{o}te d’Azur under reference number ANR-15-IDEX-01.

\end{acknowledgements}

\bibliographystyle{aa}
\bibliography{pozzi}

\begin{thebibliography}{63}
\expandafter\ifx\csname natexlab\endcsname\relax\def\natexlab#1{#1}\fi

\bibitem[{{Bagley} {et~al.}(2023){Bagley}, {Finkelstein}, {Koekemoer},
  {Ferguson}, {Arrabal Haro}, {Dickinson}, {Kartaltepe}, {Papovich},
  {P{\'e}rez-Gonz{\'a}lez}, {Pirzkal}, {Somerville}, {Willmer}, {Yang}, {Yung},
  {Fontana}, {Grazian}, {Grogin}, {Hirschmann}, {Kewley}, {Kirkpatrick},
  {Kocevski}, {Lotz}, {Medrano}, {Morales}, {Pentericci}, {Ravindranath},
  {Trump}, {Wilkins}, {Calabr{\`o}}, {Cooper}, {Costantin}, {de la Vega},
  {Hilbert}, {Hutchison}, {Larson}, {Lucas}, {McGrath}, {Ryan}, {Wang}, \&
  {Wuyts}}]{2023ApJ...946L..12B}
{Bagley}, M.~B., {Finkelstein}, S.~L., {Koekemoer}, A.~M., {et~al.} 2023,
  \apjl, 946, L12

\bibitem[{{Barro} {et~al.}(2016){Barro}, {Kriek}, {P{\'e}rez-Gonz{\'a}lez},
  {Trump}, {Koo}, {Faber}, {Dekel}, {Primack}, {Guo}, {Kocevski},
  {Mu{\~n}oz-Mateos}, {Rujopakarn}, \& {Seth}}]{2016ApJ...827L..32B}
{Barro}, G., {Kriek}, M., {P{\'e}rez-Gonz{\'a}lez}, P.~G., {et~al.} 2016,
  \apjl, 827, L32

\bibitem[{{B{\'e}thermin} {et~al.}(2023){B{\'e}thermin}, {Accard}, {Guillaume},
  {Dessauges-Zavadsky}, {Ibar}, {Cassata}, {Devereaux}, {Faisst}, {Freundlich},
  {Jones}, {Kraljic}, {Algera}, {Amor{\'\i}n}, {Bardelli}, {Boquien}, {Buat},
  {Donghia}, {Dubois}, {Ferrara}, {Fudamoto}, {Ginolfi}, {Guillard},
  {Giavalisco}, {Gruppioni}, {Gururajan}, {Hathi}, {Hayward}, {Koekemoer},
  {Lemaux}, {Magdis}, {Molina}, {Narayanan}, {Mayer}, {Pozzi}, {Rizzo},
  {Romano}, {Tasca}, {Theul{\'e}}, {Vergani}, {Vallini}, {Zamorani}, {Zanella},
  \& {Zucca}}]{2023A&A...680L...8B}
{B{\'e}thermin}, M., {Accard}, C., {Guillaume}, C., {et~al.} 2023, \aap, 680,
  L8

\bibitem[{{B{\'e}thermin} {et~al.}(2020){B{\'e}thermin}, {Fudamoto}, {Ginolfi},
  {Loiacono}, {Khusanova}, {Capak}, {Cassata}, {Faisst}, {Le F{\`e}vre},
  {Schaerer}, {Silverman}, {Yan}, {Amorin}, {Bardelli}, {Boquien}, {Cimatti},
  {Davidzon}, {Dessauges-Zavadsky}, {Fujimoto}, {Gruppioni}, {Hathi}, {Ibar},
  {Jones}, {Koekemoer}, {Lagache}, {Lemaux}, {Moreau}, {Oesch}, {Pozzi},
  {Riechers}, {Talia}, {Toft}, {Vallini}, {Vergani}, {Zamorani}, \&
  {Zucca}}]{2020A&A...643A...2B}
{B{\'e}thermin}, M., {Fudamoto}, Y., {Ginolfi}, M., {et~al.} 2020, \aap, 643,
  A2

\bibitem[{{B{\'e}thermin} {et~al.}(2017){B{\'e}thermin}, {Wu}, {Lagache},
  {Davidzon}, {Ponthieu}, {Cousin}, {Wang}, {Dor{\'e}}, {Daddi}, \&
  {Lapi}}]{2017A&A...607A..89B}
{B{\'e}thermin}, M., {Wu}, H.-Y., {Lagache}, G., {et~al.} 2017, \aap, 607, A89

\bibitem[{{Bianchi}(2013)}]{2013A&A...552A..89B}
{Bianchi}, S. 2013, \aap, 552, A89

\bibitem[{{Calistro Rivera} {et~al.}(2018){Calistro Rivera}, {Hodge}, {Smail},
  {Swinbank}, {Weiss}, {Wardlow}, {Walter}, {Rybak}, {Chen}, {Brandt},
  {Coppin}, {da Cunha}, {Dannerbauer}, {Greve}, {Karim}, {Knudsen},
  {Schinnerer}, {Simpson}, {Venemans}, \& {van der Werf}}]{2018ApJ...863...56C}
{Calistro Rivera}, G., {Hodge}, J.~A., {Smail}, I., {et~al.} 2018, \apj, 863,
  56

\bibitem[{{Calura} {et~al.}(2017){Calura}, {Pozzi}, {Cresci}, {Santini},
  {Gruppioni}, {Pozzetti}, {Gilli}, {Matteucci}, \&
  {Maiolino}}]{2017MNRAS.465...54C}
{Calura}, F., {Pozzi}, F., {Cresci}, G., {et~al.} 2017, \mnras, 465, 54

\bibitem[{{CASA Team} {et~al.}(2022){CASA Team}, {Bean}, {Bhatnagar}, {Castro},
  {Donovan Meyer}, {Emonts}, {Garcia}, {Garwood}, {Golap}, {Gonzalez Villalba},
  {Harris}, {Hayashi}, {Hoskins}, {Hsieh}, {Jagannathan}, {Kawasaki},
  {Keimpema}, {Kettenis}, {Lopez}, {Marvil}, {Masters}, {McNichols},
  {Mehringer}, {Miel}, {Moellenbrock}, {Montesino}, {Nakazato}, {Ott}, {Petry},
  {Pokorny}, {Raba}, {Rau}, {Schiebel}, {Schweighart}, {Sekhar}, {Shimada},
  {Small}, {Steeb}, {Sugimoto}, {Suoranta}, {Tsutsumi}, {van Bemmel},
  {Verkouter}, {Wells}, {Xiong}, {Szomoru}, {Griffith}, {Glendenning}, \&
  {Kern}}]{2022PASP..134k4501C}
{CASA Team}, {Bean}, B., {Bhatnagar}, S., {et~al.} 2022, \pasp, 134, 114501

\bibitem[{{Casasola} {et~al.}(2020){Casasola}, {Bianchi}, {De Vis}, {Magrini},
  {Corbelli}, {Clark}, {Fritz}, {Nersesian}, {Viaene}, {Baes}, {Cassar{\`a}},
  {Davies}, {De Looze}, {Dobbels}, {Galametz}, {Galliano}, {Jones}, {Madden},
  {Mosenkov}, {Tr{\v{c}}ka}, \& {Xilouris}}]{2020A&A...633A.100C}
{Casasola}, V., {Bianchi}, S., {De Vis}, P., {et~al.} 2020, \aap, 633, A100

\bibitem[{{Cheng} {et~al.}(2020){Cheng}, {Ibar}, {Smail}, {Molina}, {Sobral},
  {Escala}, {Best}, {Cochrane}, {Gillman}, {Swinbank}, {Ivison}, {Huang},
  {Hughes}, {Villard}, \& {Cirasuolo}}]{2020MNRAS.499.5241C}
{Cheng}, C., {Ibar}, E., {Smail}, I., {et~al.} 2020, \mnras, 499, 5241

\bibitem[{{Cochrane} {et~al.}(2019){Cochrane}, {Hayward},
  {Angl{\'e}s-Alc{\'a}zar}, {Lotz}, {Parsotan}, {Ma}, {Kere{\v{s}}},
  {Feldmann}, {Faucher-Gigu{\`e}re}, \& {Hopkins}}]{2019MNRAS.488.1779C}
{Cochrane}, R.~K., {Hayward}, C.~C., {Angl{\'e}s-Alc{\'a}zar}, D., {et~al.}
  2019, \mnras, 488, 1779

\bibitem[{{Cochrane} {et~al.}(2023){Cochrane}, {Hayward},
  {Angl{\'e}s-Alc{\'a}zar}, \& {Somerville}}]{2023MNRAS.518.5522C}
{Cochrane}, R.~K., {Hayward}, C.~C., {Angl{\'e}s-Alc{\'a}zar}, D., \&
  {Somerville}, R.~S. 2023, \mnras, 518, 5522

\bibitem[{{Cole} {et~al.}(2023){Cole}, {Papovich}, {Finkelstein}, {Bagley},
  {Dickinson}, {Iyer}, {Yung}, {Ciesla}, {Amorin}, {Arrabal Haro},
  {Bhatawdekar}, {Calabro}, {Cleri}, {de la Vega}, {Dekel}, {Endsley},
  {Gawiser}, {Giavalisco}, {Hathi}, {Hirschmann}, {Holwerda}, {Kartaltepe},
  {Koekemoer}, {Lucas}, {Mascia}, {Mobasher}, {Perez-Gonzalez}, {Rodighiero},
  {Ronayne}, {Tachhella}, {Weiner}, \& {Wilkins}}]{2023arXiv231210152C}
{Cole}, J.~W., {Papovich}, C., {Finkelstein}, S.~L., {et~al.} 2023, arXiv
  e-prints, arXiv:2312.10152

\bibitem[{{Cooke} {et~al.}(2018){Cooke}, {Smail}, {Swinbank}, {Stach}, {An},
  {Gullberg}, {Almaini}, {Simpson}, {Wardlow}, {Blain}, {Chapman}, {Chen},
  {Conselice}, {Coppin}, {Farrah}, {Maltby}, {Micha{\l}owski}, {Scott},
  {Simpson}, {Thomson}, \& {van der Werf}}]{2018ApJ...861..100C}
{Cooke}, E.~A., {Smail}, I., {Swinbank}, A.~M., {et~al.} 2018, \apj, 861, 100

\bibitem[{{Dekel} {et~al.}(2013){Dekel}, {Zolotov}, {Tweed}, {Cacciato},
  {Ceverino}, \& {Primack}}]{2013MNRAS.435..999D}
{Dekel}, A., {Zolotov}, A., {Tweed}, D., {et~al.} 2013, \mnras, 435, 999

\bibitem[{{Dessauges-Zavadsky} {et~al.}(2020){Dessauges-Zavadsky}, {Ginolfi},
  {Pozzi}, {B{\'e}thermin}, {Le F{\`e}vre}, {Fujimoto}, {Silverman}, {Jones},
  {Vallini}, {Schaerer}, {Faisst}, {Khusanova}, {Fudamoto}, {Cassata},
  {Loiacono}, {Capak}, {Yan}, {Amorin}, {Bardelli}, {Boquien}, {Cimatti},
  {Gruppioni}, {Hathi}, {Ibar}, {Koekemoer}, {Lemaux}, {Narayanan}, {Oesch},
  {Rodighiero}, {Romano}, {Talia}, {Toft}, {Vergani}, {Zamorani}, \&
  {Zucca}}]{2020A&A...643A...5D}
{Dessauges-Zavadsky}, M., {Ginolfi}, M., {Pozzi}, F., {et~al.} 2020, \aap, 643,
  A5

\bibitem[{{Devereaux} {et~al.}(2023){Devereaux}, {Cassata}, {Ibar}, {Accard},
  {Guillaume}, {B{\'e}thermin}, {Dessauges-Zavadsky}, {Faisst}, {Jones},
  {Zanella}, {Bardelli}, {Boquien}, {D'Onghia}, {Giavalisco}, {Ginolfi},
  {Gobat}, {Hayward}, {Koekemoer}, {Lemaux}, {Magdis}, {Mendez-Hernandez},
  {Molina}, {Pozzi}, {Romano}, {Tasca}, {Vergani}, {Zamorani}, \&
  {Zucca}}]{2023arXiv231111493D}
{Devereaux}, T., {Cassata}, P., {Ibar}, E., {et~al.} 2023, arXiv e-prints,
  arXiv:2311.11493

\bibitem[{{Faisst} {et~al.}(2020){Faisst}, {Schaerer}, {Lemaux}, {Oesch},
  {Fudamoto}, {Cassata}, {B{\'e}thermin}, {Capak}, {Le F{\`e}vre}, {Silverman},
  {Yan}, {Ginolfi}, {Koekemoer}, {Morselli}, {Amor{\'\i}n}, {Bardelli},
  {Boquien}, {Brammer}, {Cimatti}, {Dessauges-Zavadsky}, {Fujimoto},
  {Gruppioni}, {Hathi}, {Hemmati}, {Ibar}, {Jones}, {Khusanova}, {Loiacono},
  {Pozzi}, {Talia}, {Tasca}, {Riechers}, {Rodighiero}, {Romano}, {Scoville},
  {Toft}, {Vallini}, {Vergani}, {Zamorani}, \& {Zucca}}]{2020ApJS..247...61F}
{Faisst}, A.~L., {Schaerer}, D., {Lemaux}, B.~C., {et~al.} 2020, \apjs, 247, 61

\bibitem[{{Finkelstein} {et~al.}(2022){Finkelstein}, {Bagley}, {Arrabal Haro},
  {Dickinson}, {Ferguson}, {Kartaltepe}, {Papovich}, {Burgarella}, {Kocevski},
  {Huertas-Company}, {Iyer}, {Koekemoer}, {Larson}, {P{\'e}rez-Gonz{\'a}lez},
  {Rose}, {Tacchella}, {Wilkins}, {Chworowsky}, {Medrano}, {Morales},
  {Somerville}, {Yung}, {Fontana}, {Giavalisco}, {Grazian}, {Grogin}, {Kewley},
  {Kirkpatrick}, {Kurczynski}, {Lotz}, {Pentericci}, {Pirzkal}, {Ravindranath},
  {Ryan}, {Trump}, {Yang}, {Almaini}, {Amor{\'\i}n}, {Annunziatella},
  {Backhaus}, {Barro}, {Behroozi}, {Bell}, {Bhatawdekar}, {Bisigello}, {Bromm},
  {Buat}, {Buitrago}, {Calabr{\`o}}, {Casey}, {Castellano}, {Ch{\'a}vez Ortiz},
  {Ciesla}, {Cleri}, {Cohen}, {Cole}, {Cooke}, {Cooper}, {Cooray}, {Costantin},
  {Cox}, {Croton}, {Daddi}, {Dav{\'e}}, {de La Vega}, {Dekel}, {Elbaz},
  {Estrada-Carpenter}, {Faber}, {Fern{\'a}ndez}, {Finkelstein}, {Freundlich},
  {Fujimoto}, {Garc{\'\i}a-Argum{\'a}nez}, {Gardner}, {Gawiser},
  {G{\'o}mez-Guijarro}, {Guo}, {Hamblin}, {Hamilton}, {Hathi}, {Holwerda},
  {Hirschmann}, {Hutchison}, {Jaskot}, {Jha}, {Jogee}, {Juneau}, {Jung},
  {Kassin}, {Le Bail}, {Leung}, {Lucas}, {Magnelli}, {Mantha}, {Matharu},
  {McGrath}, {McIntosh}, {Merlin}, {Mobasher}, {Newman}, {Nicholls}, {Pandya},
  {Rafelski}, {Ronayne}, {Santini}, {Seill{\'e}}, {Shah}, {Shen}, {Simons},
  {Snyder}, {Stanway}, {Straughn}, {Teplitz}, {Vanderhoof}, {Vega-Ferrero},
  {Wang}, {Weiner}, {Willmer}, {Wuyts}, {Zavala}, \& {Ceers
  Team}}]{2022ApJ...940L..55F}
{Finkelstein}, S.~L., {Bagley}, M.~B., {Arrabal Haro}, P., {et~al.} 2022,
  \apjl, 940, L55

\bibitem[{{Franco} {et~al.}(2020){Franco}, {Elbaz}, {Zhou}, {Magnelli},
  {Schreiber}, {Ciesla}, {Dickinson}, {Nagar}, {Magdis}, {Alexander},
  {B{\'e}thermin}, {Demarco}, {Daddi}, {Wang}, {Mullaney}, {Sargent}, {Inami},
  {Shu}, {Bournaud}, {Chary}, {Coogan}, {Ferguson}, {Finkelstein},
  {Giavalisco}, {G{\'o}mez-Guijarro}, {Iono}, {Juneau}, {Lagache}, {Lin},
  {Motohara}, {Okumura}, {Pannella}, {Papovich}, {Pope}, {Rujopakarn},
  {Silverman}, \& {Xiao}}]{2020A&A...643A..30F}
{Franco}, M., {Elbaz}, D., {Zhou}, L., {et~al.} 2020, \aap, 643, A30

\bibitem[{{Fujimoto} {et~al.}(2017){Fujimoto}, {Ouchi}, {Shibuya}, \&
  {Nagai}}]{2017ApJ...850...83F}
{Fujimoto}, S., {Ouchi}, M., {Shibuya}, T., \& {Nagai}, H. 2017, \apj, 850, 83

\bibitem[{{Fujimoto} {et~al.}(2020){Fujimoto}, {Silverman}, {Bethermin},
  {Ginolfi}, {Jones}, {Le F{\`e}vre}, {Dessauges-Zavadsky}, {Rujopakarn},
  {Faisst}, {Fudamoto}, {Cassata}, {Morselli}, {Maiolino}, {Schaerer}, {Capak},
  {Yan}, {Vallini}, {Toft}, {Loiacono}, {Zamorani}, {Talia}, {Narayanan},
  {Hathi}, {Lemaux}, {Boquien}, {Amorin}, {Ibar}, {Koekemoer},
  {M{\'e}ndez-Hern{\'a}ndez}, {Bardelli}, {Vergani}, {Zucca}, {Romano}, \&
  {Cimatti}}]{2020ApJ...900....1F}
{Fujimoto}, S., {Silverman}, J.~D., {Bethermin}, M., {et~al.} 2020, \apj, 900,
  1

\bibitem[{{Giacconi} {et~al.}(2002){Giacconi}, {Zirm}, {Wang}, {Rosati},
  {Nonino}, {Tozzi}, {Gilli}, {Mainieri}, {Hasinger}, {Kewley}, {Bergeron},
  {Borgani}, {Gilmozzi}, {Grogin}, {Koekemoer}, {Schreier}, {Zheng}, \&
  {Norman}}]{2002ApJS..139..369G}
{Giacconi}, R., {Zirm}, A., {Wang}, J., {et~al.} 2002, \apjs, 139, 369

\bibitem[{{Gilli} {et~al.}(2022){Gilli}, {Norman}, {Calura}, {Vito}, {Decarli},
  {Marchesi}, {Iwasawa}, {Comastri}, {Lanzuisi}, {Pozzi}, {D'Amato}, {Vignali},
  {Brusa}, {Mignoli}, \& {Cox}}]{2022A&A...666A..17G}
{Gilli}, R., {Norman}, C., {Calura}, F., {et~al.} 2022, \aap, 666, A17

\bibitem[{{G{\'o}mez-Guijarro}
  {et~al.}(2022{\natexlab{a}}){G{\'o}mez-Guijarro}, {Elbaz}, {Xiao},
  {B{\'e}thermin}, {Franco}, {Magnelli}, {Daddi}, {Dickinson}, {Demarco},
  {Inami}, {Rujopakarn}, {Magdis}, {Shu}, {Chary}, {Zhou}, {Alexander},
  {Bournaud}, {Ciesla}, {Ferguson}, {Finkelstein}, {Giavalisco}, {Iono},
  {Juneau}, {Kartaltepe}, {Lagache}, {Le Floc'h}, {Leiton}, {Lin}, {Motohara},
  {Mullaney}, {Okumura}, {Pannella}, {Papovich}, {Pope}, {Sargent},
  {Silverman}, {Treister}, \& {Wang}}]{2022A&A...658A..43G}
{G{\'o}mez-Guijarro}, C., {Elbaz}, D., {Xiao}, M., {et~al.} 2022{\natexlab{a}},
  \aap, 658, A43

\bibitem[{{G{\'o}mez-Guijarro}
  {et~al.}(2022{\natexlab{b}}){G{\'o}mez-Guijarro}, {Elbaz}, {Xiao}, {Kokorev},
  {Magdis}, {Magnelli}, {Daddi}, {Valentino}, {Sargent}, {Dickinson},
  {B{\'e}thermin}, {Franco}, {Pope}, {Kalita}, {Ciesla}, {Demarco}, {Inami},
  {Rujopakarn}, {Shu}, {Wang}, {Zhou}, {Alexander}, {Bournaud}, {Chary},
  {Ferguson}, {Finkelstein}, {Giavalisco}, {Iono}, {Juneau}, {Kartaltepe},
  {Lagache}, {Le Floc'h}, {Leiton}, {Leroy}, {Lin}, {Motohara}, {Mullaney},
  {Okumura}, {Pannella}, {Papovich}, \& {Treister}}]{2022A&A...659A.196G}
{G{\'o}mez-Guijarro}, C., {Elbaz}, D., {Xiao}, M., {et~al.} 2022{\natexlab{b}},
  \aap, 659, A196

\bibitem[{{Herrera-Camus} {et~al.}(2018){Herrera-Camus}, {Sturm},
  {Graci{\'a}-Carpio}, {Lutz}, {Contursi}, {Veilleux}, {Fischer},
  {Gonz{\'a}lez-Alfonso}, {Poglitsch}, {Tacconi}, {Genzel}, {Maiolino},
  {Sternberg}, {Davies}, \& {Verma}}]{2018ApJ...861...95H}
{Herrera-Camus}, R., {Sturm}, E., {Graci{\'a}-Carpio}, J., {et~al.} 2018, \apj,
  861, 95

\bibitem[{{Hodge} \& {da Cunha}(2020)}]{2020RSOS....700556H}
{Hodge}, J.~A. \& {da Cunha}, E. 2020, Royal Society Open Science, 7, 200556

\bibitem[{{Hopkins} {et~al.}(2018){Hopkins}, {Wetzel}, {Kere{\v{s}}},
  {Faucher-Gigu{\`e}re}, {Quataert}, {Boylan-Kolchin}, {Murray}, {Hayward},
  {Garrison-Kimmel}, {Hummels}, {Feldmann}, {Torrey}, {Ma},
  {Angl{\'e}s-Alc{\'a}zar}, {Su}, {Orr}, {Schmitz}, {Escala}, {Sanderson},
  {Grudi{\'c}}, {Hafen}, {Kim}, {Fitts}, {Bullock}, {Wheeler}, {Chan},
  {Elbert}, \& {Narayanan}}]{2018MNRAS.480..800H}
{Hopkins}, P.~F., {Wetzel}, A., {Kere{\v{s}}}, D., {et~al.} 2018, \mnras, 480,
  800

\bibitem[{{Jin} {et~al.}(2018){Jin}, {Daddi}, {Liu}, {Smol{\v{c}}i{\'c}},
  {Schinnerer}, {Calabr{\`o}}, {Gu}, {Delhaize}, {Delvecchio}, {Gao},
  {Salvato}, {Puglisi}, {Dickinson}, {Bertoldi}, {Sargent}, {Novak}, {Magdis},
  {Aretxaga}, {Wilson}, \& {Capak}}]{2018ApJ...864...56J}
{Jin}, S., {Daddi}, E., {Liu}, D., {et~al.} 2018, \apj, 864, 56

\bibitem[{{Jin} {et~al.}(2022){Jin}, {Daddi}, {Magdis}, {Liu}, {Weaver}, {Tan},
  {Valentino}, {Gao}, {Schinnerer}, {Calabr{\`o}}, {Gu}, \&
  {Sese}}]{2022A&A...665A...3J}
{Jin}, S., {Daddi}, E., {Magdis}, G.~E., {et~al.} 2022, \aap, 665, A3

\bibitem[{{Jones} {et~al.}(2020){Jones}, {B{\'e}thermin}, {Fudamoto},
  {Ginolfi}, {Capak}, {Cassata}, {Faisst}, {Le F{\`e}vre}, {Schaerer},
  {Silverman}, {Yan}, {Bardelli}, {Boquien}, {Cimatti}, {Dessauges-Zavadsky},
  {Giavalisco}, {Gruppioni}, {Ibar}, {Khusanova}, {Koekemoer}, {Lemaux},
  {Loiacono}, {Maiolino}, {Oesch}, {Pozzi}, {Riechers}, {Rodighiero}, {Talia},
  {Vallini}, {Vergani}, {Zamorani}, \& {Zucca}}]{2020MNRAS.491L..18J}
{Jones}, G.~C., {B{\'e}thermin}, M., {Fudamoto}, Y., {et~al.} 2020, \mnras,
  491, L18

\bibitem[{{Koekemoer} {et~al.}(2007){Koekemoer}, {Aussel}, {Calzetti}, {Capak},
  {Giavalisco}, {Kneib}, {Leauthaud}, {Le F{\`e}vre}, {McCracken}, {Massey},
  {Mobasher}, {Rhodes}, {Scoville}, \& {Shopbell}}]{2007ApJS..172..196K}
{Koekemoer}, A.~M., {Aussel}, H., {Calzetti}, D., {et~al.} 2007, \apjs, 172,
  196

\bibitem[{{Koekemoer} {et~al.}(2011){Koekemoer}, {Faber}, {Ferguson}, {Grogin},
  {Kocevski}, {Koo}, {Lai}, {Lotz}, {Lucas}, {McGrath}, {Ogaz}, {Rajan},
  {Riess}, {Rodney}, {Strolger}, {Casertano}, {Castellano}, {Dahlen},
  {Dickinson}, {Dolch}, {Fontana}, {Giavalisco}, {Grazian}, {Guo}, {Hathi},
  {Huang}, {van der Wel}, {Yan}, {Acquaviva}, {Alexander}, {Almaini}, {Ashby},
  {Barden}, {Bell}, {Bournaud}, {Brown}, {Caputi}, {Cassata}, {Challis},
  {Chary}, {Cheung}, {Cirasuolo}, {Conselice}, {Roshan Cooray}, {Croton},
  {Daddi}, {Dav{\'e}}, {de Mello}, {de Ravel}, {Dekel}, {Donley}, {Dunlop},
  {Dutton}, {Elbaz}, {Fazio}, {Filippenko}, {Finkelstein}, {Frazer}, {Gardner},
  {Garnavich}, {Gawiser}, {Gruetzbauch}, {Hartley}, {H{\"a}ussler},
  {Herrington}, {Hopkins}, {Huang}, {Jha}, {Johnson}, {Kartaltepe},
  {Khostovan}, {Kirshner}, {Lani}, {Lee}, {Li}, {Madau}, {McCarthy},
  {McIntosh}, {McLure}, {McPartland}, {Mobasher}, {Moreira}, {Mortlock},
  {Moustakas}, {Mozena}, {Nandra}, {Newman}, {Nielsen}, {Niemi}, {Noeske},
  {Papovich}, {Pentericci}, {Pope}, {Primack}, {Ravindranath}, {Reddy},
  {Renzini}, {Rix}, {Robaina}, {Rosario}, {Rosati}, {Salimbeni}, {Scarlata},
  {Siana}, {Simard}, {Smidt}, {Snyder}, {Somerville}, {Spinrad}, {Straughn},
  {Telford}, {Teplitz}, {Trump}, {Vargas}, {Villforth}, {Wagner}, {Wandro},
  {Wechsler}, {Weiner}, {Wiklind}, {Wild}, {Wilson}, {Wuyts}, \&
  {Yun}}]{2011ApJS..197...36K}
{Koekemoer}, A.~M., {Faber}, S.~M., {Ferguson}, H.~C., {et~al.} 2011, \apjs,
  197, 36

\bibitem[{{Le F{\`e}vre} {et~al.}(2020){Le F{\`e}vre}, {B{\'e}thermin},
  {Faisst}, {Jones}, {Capak}, {Cassata}, {Silverman}, {Schaerer}, {Yan},
  {Amorin}, {Bardelli}, {Boquien}, {Cimatti}, {Dessauges-Zavadsky},
  {Giavalisco}, {Hathi}, {Fudamoto}, {Fujimoto}, {Ginolfi}, {Gruppioni},
  {Hemmati}, {Ibar}, {Koekemoer}, {Khusanova}, {Lagache}, {Lemaux}, {Loiacono},
  {Maiolino}, {Mancini}, {Narayanan}, {Morselli}, {M{\'e}ndez-Hern{\`a}ndez},
  {Oesch}, {Pozzi}, {Romano}, {Riechers}, {Scoville}, {Talia}, {Tasca},
  {Thomas}, {Toft}, {Vallini}, {Vergani}, {Walter}, {Zamorani}, \&
  {Zucca}}]{2020A&A...643A...1L}
{Le F{\`e}vre}, O., {B{\'e}thermin}, M., {Faisst}, A., {et~al.} 2020, \aap,
  643, A1

\bibitem[{{Liu} {et~al.}(2018){Liu}, {Daddi}, {Dickinson}, {Owen}, {Pannella},
  {Sargent}, {B{\'e}thermin}, {Magdis}, {Gao}, {Shu}, {Wang}, {Jin}, \&
  {Inami}}]{2018ApJ...853..172L}
{Liu}, D., {Daddi}, E., {Dickinson}, M., {et~al.} 2018, \apj, 853, 172

\bibitem[{{Marinacci} {et~al.}(2018){Marinacci}, {Vogelsberger}, {Pakmor},
  {Torrey}, {Springel}, {Hernquist}, {Nelson}, {Weinberger}, {Pillepich},
  {Naiman}, \& {Genel}}]{2018MNRAS.480.5113M}
{Marinacci}, F., {Vogelsberger}, M., {Pakmor}, R., {et~al.} 2018, \mnras, 480,
  5113

\bibitem[{{Mart{\'\i}-Vidal} {et~al.}(2012){Mart{\'\i}-Vidal},
  {P{\'e}rez-Torres}, \& {Lobanov}}]{2012A&A...541A.135M}
{Mart{\'\i}-Vidal}, I., {P{\'e}rez-Torres}, M.~A., \& {Lobanov}, A.~P. 2012,
  \aap, 541, A135

\bibitem[{{Mitsuhashi} {et~al.}(2023){Mitsuhashi}, {Tadaki}, {Ikeda},
  {Herrera-Camus}, {Aravena}, {De Looze}, {F{\"o}rster Schreiber},
  {Gonz{\'a}lez-L{\'o}pez}, {Spilker}, {Assef}, {Bouwens}, {Barcos-Munoz},
  {Birkin}, {Bowler}, {Calistro Rivera}, {Davies}, {Da Cunha},
  {D{\'\i}az-Santos}, {Ferrara}, {Fisher}, {Lee}, {Li}, {Lutz}, {Rela{\~n}o},
  {Naab}, {Palla}, {Posses}, {Solimano}, {Tacconi}, {{\"U}bler}, {van der
  Giessen}, \& {Veilleux}}]{2023arXiv231117671M}
{Mitsuhashi}, I., {Tadaki}, K.-i., {Ikeda}, R., {et~al.} 2023, arXiv e-prints,
  arXiv:2311.17671

\bibitem[{{Nelson} {et~al.}(2019){Nelson}, {Pillepich}, {Springel}, {Pakmor},
  {Weinberger}, {Genel}, {Torrey}, {Vogelsberger}, {Marinacci}, \&
  {Hernquist}}]{2019MNRAS.490.3234N}
{Nelson}, D., {Pillepich}, A., {Springel}, V., {et~al.} 2019, \mnras, 490, 3234

\bibitem[{{Pastrav}(2020)}]{2020MNRAS.493.3580P}
{Pastrav}, B.~A. 2020, \mnras, 493, 3580

\bibitem[{{Peng} {et~al.}(2010){Peng}, {Ho}, {Impey}, \&
  {Rix}}]{2010AJ....139.2097P}
{Peng}, C.~Y., {Ho}, L.~C., {Impey}, C.~D., \& {Rix}, H.-W. 2010, \aj, 139,
  2097

\bibitem[{{Pillepich} {et~al.}(2019){Pillepich}, {Nelson}, {Springel},
  {Pakmor}, {Torrey}, {Weinberger}, {Vogelsberger}, {Marinacci}, {Genel}, {van
  der Wel}, \& {Hernquist}}]{2019MNRAS.490.3196P}
{Pillepich}, A., {Nelson}, D., {Springel}, V., {et~al.} 2019, \mnras, 490, 3196

\bibitem[{{Popping} {et~al.}(2022){Popping}, {Pillepich}, {Calistro Rivera},
  {Schulz}, {Hernquist}, {Kaasinen}, {Marinacci}, {Nelson}, \&
  {Vogelsberger}}]{2022MNRAS.510.3321P}
{Popping}, G., {Pillepich}, A., {Calistro Rivera}, G., {et~al.} 2022, \mnras,
  510, 3321

\bibitem[{{Pozzi} {et~al.}(2021){Pozzi}, {Calura}, {Fudamoto},
  {Dessauges-Zavadsky}, {Gruppioni}, {Talia}, {Zamorani}, {Bethermin},
  {Cimatti}, {Enia}, {Khusanova}, {Decarli}, {Le F{\`e}vre}, {Capak},
  {Cassata}, {Faisst}, {Yan}, {Schaerer}, {Silverman}, {Bardelli}, {Boquien},
  {Enia}, {Narayanan}, {Ginolfi}, {Hathi}, {Jones}, {Koekemoer}, {Lemaux},
  {Loiacono}, {Maiolino}, {Riechers}, {Rodighiero}, {Romano}, {Vallini},
  {Vergani}, \& {Zucca}}]{2021A&A...653A..84P}
{Pozzi}, F., {Calura}, F., {Fudamoto}, Y., {et~al.} 2021, \aap, 653, A84

\bibitem[{{Puglisi} {et~al.}(2021){Puglisi}, {Daddi}, {Valentino}, {Magdis},
  {Liu}, {Kokorev}, {Circosta}, {Elbaz}, {Bournaud}, {Gomez-Guijarro}, {Jin},
  {Madden}, {Sargent}, \& {Swinbank}}]{2021MNRAS.508.5217P}
{Puglisi}, A., {Daddi}, E., {Valentino}, F., {et~al.} 2021, \mnras, 508, 5217

\bibitem[{{Rujopakarn} {et~al.}(2016){Rujopakarn}, {Dunlop}, {Rieke}, {Ivison},
  {Cibinel}, {Nyland}, {Jagannathan}, {Silverman}, {Alexander}, {Biggs},
  {Bhatnagar}, {Ballantyne}, {Dickinson}, {Elbaz}, {Geach}, {Hayward},
  {Kirkpatrick}, {McLure}, {Micha{\l}owski}, {Miller}, {Narayanan}, {Owen},
  {Pannella}, {Papovich}, {Pope}, {Rau}, {Robertson}, {Scott}, {Swinbank}, {van
  der Werf}, {van Kampen}, {Weiner}, \& {Windhorst}}]{2016ApJ...833...12R}
{Rujopakarn}, W., {Dunlop}, J.~S., {Rieke}, G.~H., {et~al.} 2016, \apj, 833, 12

\bibitem[{{Schaerer} {et~al.}(2020){Schaerer}, {Ginolfi}, {B{\'e}thermin},
  {Fudamoto}, {Oesch}, {Le F{\`e}vre}, {Faisst}, {Capak}, {Cassata},
  {Silverman}, {Yan}, {Jones}, {Amorin}, {Bardelli}, {Boquien}, {Cimatti},
  {Dessauges-Zavadsky}, {Giavalisco}, {Hathi}, {Fujimoto}, {Ibar}, {Koekemoer},
  {Lagache}, {Lemaux}, {Loiacono}, {Maiolino}, {Narayanan}, {Morselli},
  {M{\'e}ndez-Hern{\`a}ndez}, {Pozzi}, {Riechers}, {Talia}, {Toft}, {Vallini},
  {Vergani}, {Zamorani}, \& {Zucca}}]{2020A&A...643A...3S}
{Schaerer}, D., {Ginolfi}, M., {B{\'e}thermin}, M., {et~al.} 2020, \aap, 643,
  A3

\bibitem[{{Scoville} {et~al.}(2007){Scoville}, {Aussel}, {Brusa}, {Capak},
  {Carollo}, {Elvis}, {Giavalisco}, {Guzzo}, {Hasinger}, {Impey}, {Kneib},
  {LeFevre}, {Lilly}, {Mobasher}, {Renzini}, {Rich}, {Sanders}, {Schinnerer},
  {Schminovich}, {Shopbell}, {Taniguchi}, \& {Tyson}}]{2007ApJS..172....1S}
{Scoville}, N., {Aussel}, H., {Brusa}, M., {et~al.} 2007, \apjs, 172, 1

\bibitem[{{Scoville} {et~al.}(2023){Scoville}, {Faisst}, {Weaver}, {Toft},
  {McCracken}, {Ilbert}, {Diaz-Santos}, {Staguhn}, {Koda}, {Casey}, {Sanders},
  {Mobasher}, {Chartab}, {Sattari}, {Capak}, {Vanden Bout}, {Bongiorno},
  {Vlahakis}, {Sheth}, {Yun}, {Aussel}, {Laigle}, \&
  {Masters}}]{2023ApJ...943...82S}
{Scoville}, N., {Faisst}, A., {Weaver}, J., {et~al.} 2023, \apj, 943, 82

\bibitem[{{Silverman} {et~al.}(2015){Silverman}, {Daddi}, {Rodighiero},
  {Rujopakarn}, {Sargent}, {Renzini}, {Liu}, {Feruglio}, {Kashino}, {Sanders},
  {Kartaltepe}, {Nagao}, {Arimoto}, {Berta}, {B{\'e}thermin}, {Koekemoer},
  {Lutz}, {Magdis}, {Mancini}, {Onodera}, \& {Zamorani}}]{2015ApJ...812L..23S}
{Silverman}, J.~D., {Daddi}, E., {Rodighiero}, G., {et~al.} 2015, \apjl, 812,
  L23

\bibitem[{{Sommovigo} {et~al.}(2021){Sommovigo}, {Ferrara}, {Carniani},
  {Zanella}, {Pallottini}, {Gallerani}, \& {Vallini}}]{2021MNRAS.503.4878S}
{Sommovigo}, L., {Ferrara}, A., {Carniani}, S., {et~al.} 2021, \mnras, 503,
  4878

\bibitem[{{Speagle} {et~al.}(2014){Speagle}, {Steinhardt}, {Capak}, \&
  {Silverman}}]{2014ApJS..214...15S}
{Speagle}, J.~S., {Steinhardt}, C.~L., {Capak}, P.~L., \& {Silverman}, J.~D.
  2014, \apjs, 214, 15

\bibitem[{{Springel} {et~al.}(2018){Springel}, {Pakmor}, {Pillepich},
  {Weinberger}, {Nelson}, {Hernquist}, {Vogelsberger}, {Genel}, {Torrey},
  {Marinacci}, \& {Naiman}}]{2018MNRAS.475..676S}
{Springel}, V., {Pakmor}, R., {Pillepich}, A., {et~al.} 2018, \mnras, 475, 676

\bibitem[{{Tadaki} {et~al.}(2017){Tadaki}, {Genzel}, {Kodama}, {Wuyts},
  {Wisnioski}, {F{\"o}rster Schreiber}, {Burkert}, {Lang}, {Tacconi}, {Lutz},
  {Belli}, {Davies}, {Hatsukade}, {Hayashi}, {Herrera-Camus}, {Ikarashi},
  {Inoue}, {Kohno}, {Koyama}, {Mendel}, {Nakanishi}, {Shimakawa}, {Suzuki},
  {Tamura}, {Tanaka}, {{\"U}bler}, \& {Wilman}}]{2017ApJ...834..135T}
{Tadaki}, K.-i., {Genzel}, R., {Kodama}, T., {et~al.} 2017, \apj, 834, 135

\bibitem[{{Talia} {et~al.}(2018){Talia}, {Pozzi}, {Vallini}, {Cimatti},
  {Cassata}, {Fraternali}, {Brusa}, {Daddi}, {Delvecchio}, {Ibar}, {Liuzzo},
  {Vignali}, {Massardi}, {Zamorani}, {Gruppioni}, {Renzini}, {Mignoli},
  {Pozzetti}, \& {Rodighiero}}]{2018MNRAS.476.3956T}
{Talia}, M., {Pozzi}, F., {Vallini}, L., {et~al.} 2018, \mnras, 476, 3956

\bibitem[{{Valentino} {et~al.}(2020){Valentino}, {Magdis}, {Daddi}, {Liu},
  {Aravena}, {Bournaud}, {Cortzen}, {Gao}, {Jin}, {Juneau}, {Kartaltepe},
  {Kokorev}, {Lee}, {Madden}, {Narayanan}, {Popping}, \&
  {Puglisi}}]{2020ApJ...890...24V}
{Valentino}, F., {Magdis}, G.~E., {Daddi}, E., {et~al.} 2020, \apj, 890, 24

\bibitem[{{Vallini} {et~al.}(2024){Vallini}, {Witstok}, {Sommovigo},
  {Pallottini}, {Ferrara}, {Carniani}, {Kohandel}, {Smit}, {Gallerani}, \&
  {Gruppioni}}]{2024MNRAS.527...10V}
{Vallini}, L., {Witstok}, J., {Sommovigo}, L., {et~al.} 2024, \mnras, 527, 10

\bibitem[{{van der Wel} {et~al.}(2014){van der Wel}, {Franx}, {van Dokkum},
  {Skelton}, {Momcheva}, {Whitaker}, {Brammer}, {Bell}, {Rix}, {Wuyts},
  {Ferguson}, {Holden}, {Barro}, {Koekemoer}, {Chang}, {McGrath},
  {H{\"a}ussler}, {Dekel}, {Behroozi}, {Fumagalli}, {Leja}, {Lundgren},
  {Maseda}, {Nelson}, {Wake}, {Patel}, {Labb{\'e}}, {Faber}, {Grogin}, \&
  {Kocevski}}]{2014ApJ...788...28V}
{van der Wel}, A., {Franx}, M., {van Dokkum}, P.~G., {et~al.} 2014, \apj, 788,
  28

\bibitem[{{Voigt} \& {Bridle}(2010)}]{2010MNRAS.404..458V}
{Voigt}, L.~M. \& {Bridle}, S.~L. 2010, {Limitations of model-fitting methods
  for lensing shear estimation}

\bibitem[{{Ward} {et~al.}(2023){Ward}, {de la Vega}, {Mobasher}, {McGrath},
  {Iyer}, {Calabro}, {Costantin}, {Dickinson}, {Holwerda}, {Huertas-Company},
  {Hirschmann}, {Lucas}, {Pandya}, {Wilkins}, {Yung}, {Arrabal Haro}, {Bagley},
  {Finkelstein}, {Kartaltepe}, {Koekemoer}, {Papovich}, \&
  {Pirzkal}}]{2023arXiv231102162W}
{Ward}, E.~M., {de la Vega}, A., {Mobasher}, B., {et~al.} 2023, arXiv e-prints,
  arXiv:2311.02162

\bibitem[{{Witstok} {et~al.}(2022){Witstok}, {Smit}, {Maiolino}, {Kumari},
  {Aravena}, {Boogaard}, {Bouwens}, {Carniani}, {Hodge}, {Jones}, {Stefanon},
  {van der Werf}, \& {Schouws}}]{2022MNRAS.515.1751W}
{Witstok}, J., {Smit}, R., {Maiolino}, R., {et~al.} 2022, \mnras, 515, 1751

\end{thebibliography}

\end{document}